# Excitation Amplitude Dependence of Low Frequency Magneto-impedance of Amorphous $Fe_{73.5}Nb_3Cu_1Si_{13.5}B_9$ Ribbon.

**B.Kaviraj and S.K.Ghatak**

*Department of Physics and Meteorology, Indian Institute of Technology, Kharagpur, India*

*Email*: bhaskar@phy.iitkgp.ernet.in

## Abstract

Magneto-impedance (MI) of soft ferromagnetic materials is a sensitive function of the amplitude and frequency of exciting a.c magnetic field. MI (Z) of amorphous ferromagnetic ribbon of nominal composition $Fe_{73.5}Nb_3\ Cu_1\ Si_{13.5}\ B_9$ is measured at different excitation currents with frequency ranging from 30kHz to 120KHz The excitation (ac) and biasing (dc) magnetic fields $H_{dc}$ are parallel to ribbon axis. At zero dc fields, Z exhibits non-linear dependence on excitation amplitude whereas at higher d.c fields (9Oe) Z is nearly independent of excitation amplitude. The impedance is maximum at zero bias field and sharply decreases as $H_{dc}$ increases and almost zero hysterisis has been observed as $H_{dc}$ is scanned. The maximum relative change of impedance (Z(H)-Z(H=0))/Z(H=0) is found to increase from nearly 45 to 60 percent when the excitation fields goes up from 14A/m to 140A/m. The large MI is associated with screening of electromagnetic field by magnetization induced by exciting ac field.

**Keywords:** *Magneto-impedance, Ferromagnet, Magnetization.*

# Introduction

Giant Magneto-impedance (GMI) consists of a large change of total impedance of a soft magnetic conductor under the application of a static magnetic field $H_{dc}$[1]. The GMI effect appears to be the result of combined influence of screening effect of conduction electrons and softness of ferromagnetic conductor. Much of this phenomenon can be understood in the framework of classical electrodynamics [2-4]. At low frequencies, the change in magnetization generates an additional inductive voltage, $V_L$ across the conductor but at high excitation frequencies, the skin depth becomes comparable to the lower dimension of the sample and the current carrying capacity of the sample is severely impeded. The large change in magneto-impedance is associated with the reduction of magnetic response due to the biased field [5-6]. The extent of decrease of magneto-impedance for small dc fields depends on the magnetic anisotropy, which in turn is related to the domain structure of the ferromagnetic system. In ferromagnetic materials, the magnetic permeability depends upon amplitude and frequency of excitation fields along with several other parameters. In case of low excitation frequencies, the essential field dependence of ac voltage results from only the inductive part $V_L$ and is proportional to the magnetic permeability. The GMI effect depends upon current, which arises from variation of induced magnetization, which in turn depends upon configuration of exciting (a.c) and biasing fields with reference to the anisotropy field of the material. In this present communication, we report large changes of magneto-impedance as a function of amplitudes of excitation fields for two frequencies.

## Experimental

In this present experiment we used amorphous ribbon with nominal composition $Fe_{73.5}Nb_3Cu_1Si_{13.5}B_9$ produced by melt-spinning technique. The sample dimensions were 8mmx3mmx30μm. The sample was placed inside a coil that created an ac magnetic field ($h_{ac}$) parallel to length of the sample. The resistive and inductive component of MI was obtained from measurement of signal induced in a secondary wound over sample. The signal voltage was measured with the help of a Lock-in Amplifier (EG&G Model 5209) at two different excitation frequencies of 30KHz and 120KHz. The bias magnetic field $H_{dc}$ was produced by a solenoid of sensitivity of 11Oe/A.

## Results and Discussions

The relative change [$\Delta Z/Z(H=0)$] of MI of above ribbon sample is shown in Fig.1 for different amplitudes of exciting field at 30KHz frequency. Here Z(H=0) is the impedance at zero bias field $H_{dc}$. Impedance is maximum at zero dc field $H_{dc}$ and decreases with increase in $H_{dc}$.

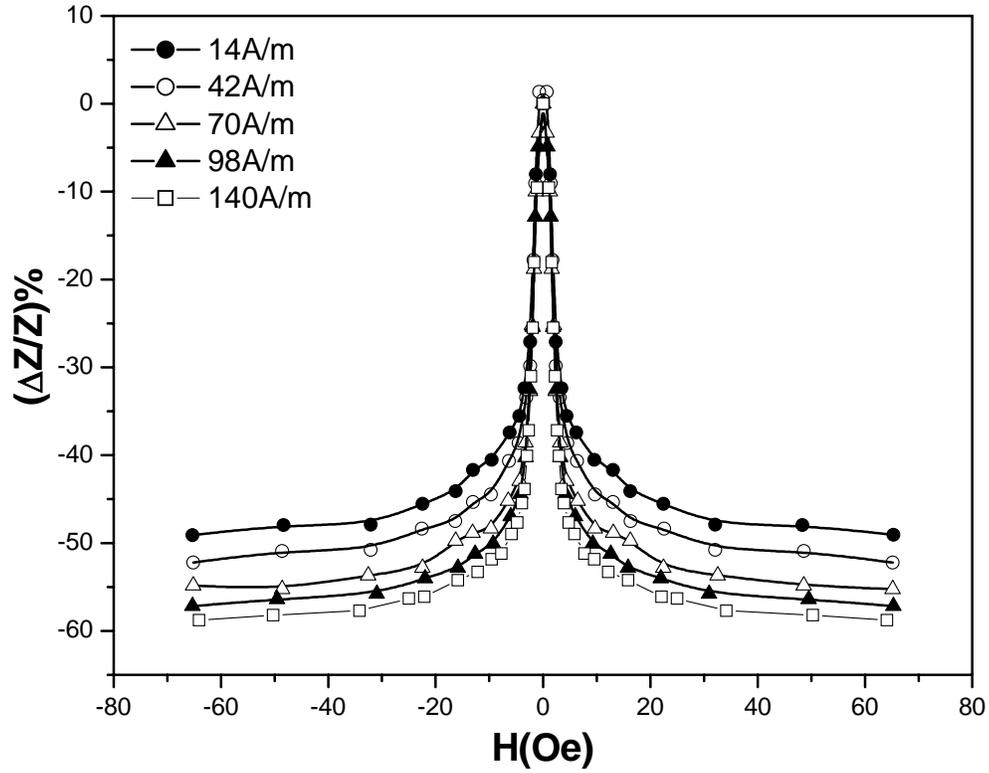

**Fig. 1**. **Field dependence of variation of ΔZ/Z (%) at various excitation ac fields.**

The MI is negative and approaches to its saturating value at field ≥ 50Oe. In presence of $H_{dc}$ the relative change of MI becomes higher for higher amplitude of exciting field. As $h_{ac}$ goes up from 14 A/m to 140 A/m, the maximum relative change in GMI is found to increase from almost 45% to 60%. This is mainly due to the higher values of Z (H=0) at larger excitation amplitude (Fig. 2).

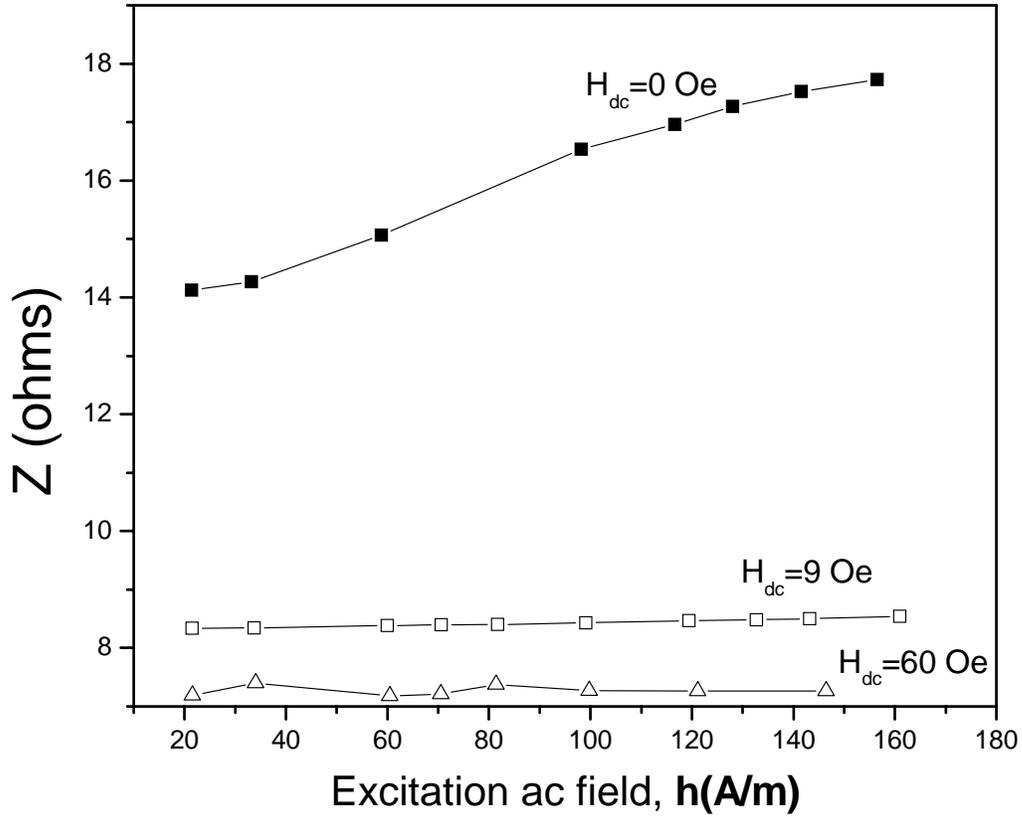

**Fig. 2**. **Dependence of Z on ac excitation fields at various biasing fields.**

In presence of higher $H_{dc}$ impedance is nearly independent of amplitude of a.c. field. Fig.2 shows that the value of MI increases non-linearly with the excitation field. Fig.3 displays impedance variation at two different frequencies. Note sharper fall of Z with increase in $H_{dc}$ at higher frequency.

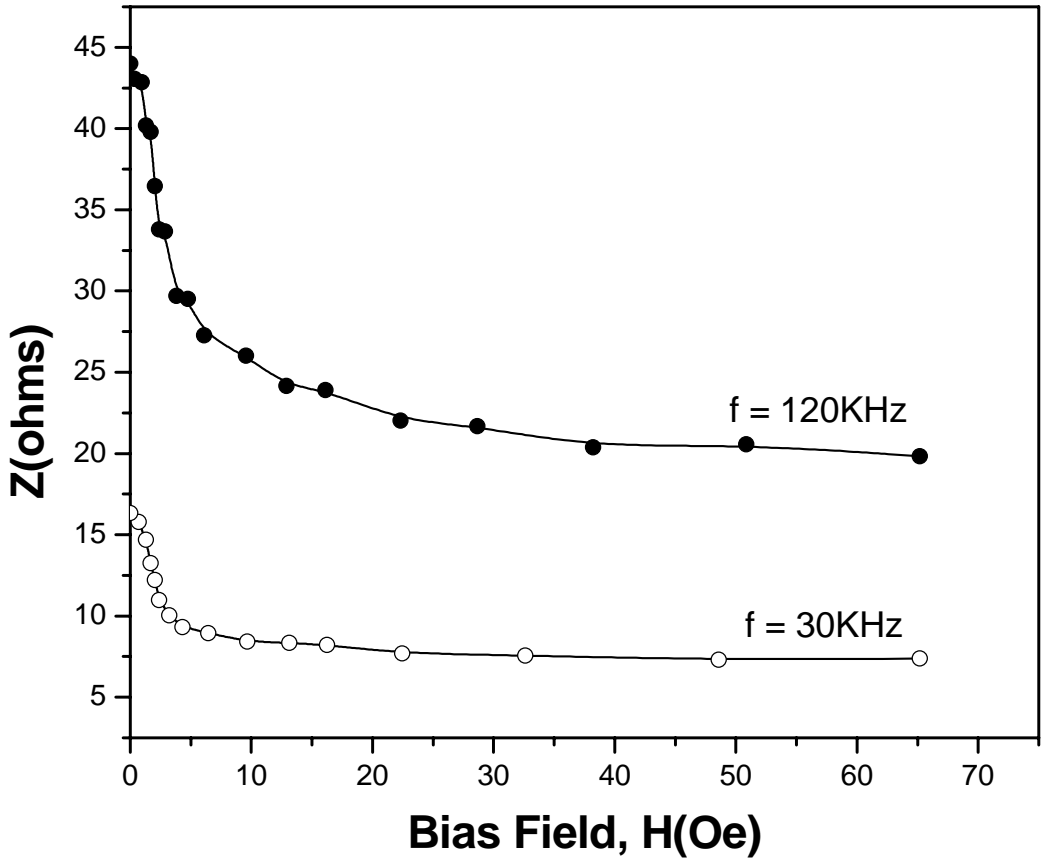

**Fig. 3. Dependence of Z on magnetic fields at different excitation frequencies.**

The impedance of a conductor is associated with the screening of electromagnetic field and in ferromagnetic conductor additional screening effect arises due to magnetization current. This current depends strongly on the magnetic response (permeability) which in turn depends on amplitude and frequency of the exciting a.c field. In soft ferromagnetic material, the response can be modulated by biasing field. In the absence of dc field, multiple domain character of the ferromagnet produces a large screening current caused by large magnetic response. Therefore, the effective skin depth is much reduced and hence current flow is strongly impeded. As the magnetic response is higher for higher value of $h_{ac}$ ($h_{ac}$ < coercive field), larger impedance results at higher $h_{ac}$. On the other hand the system becomes mono-domain in nature at large biasing field and the response is drastically reduced. This increases effective skin depth and consequent diminution of impedance. The non-linear dependence of impedance on $h_{ac}$ in multi-domain system is directly connected with the nonlinear field dependence of induced magnetization. In saturated state at large $H_{dc}$, the induced magnetization varies linearly with small $h_{ac}$ and therefore the response is independent of $h_{ac}$. This in turn leads to nearly independent behavior of Z with $h_{ac}$ (Fig. 2).

**Conclusions**

Magneto-impedance (MI) of soft ferromagnetic materials is a sensitive function of the amplitude and frequency of exciting a.c magnetic field. MI (Z) of amorphous ferromagnetic ribbon of nominal composition $Fe_{73.5}Nb_3 Cu_1 Si_{13.5} B_9$ is measured at different excitation currents with frequency ranging from 30kHz to 120KHz The excitation (ac) and biasing (dc) magnetic fields $H_{dc}$ are parallel to ribbon axis. At zero dc fields, Z exhibits non-linear dependence on excitation amplitude whereas at higher d.c fields (9Oe) Z is nearly independent of excitation amplitude. The impedance is maximum at zero bias field and sharply decreases as $H_{dc}$ increases and almost zero hysterisis has been observed as $H_{dc}$ is scanned. The maximum relative change of impedance (Z(H)-Z(H=0))/Z(H=0) is found to increase from nearly 45 to 60 percent when the excitation fields goes up from 14A/m to 140A/m. The large MI is associated with screening of electromagnetic field by magnetization induced by exciting ac field.